\documentclass[journal,onecolumn,12pt]{IEEEtran}
\renewcommand{\baselinestretch}{1.5}
\newcommand{\bega}{\begin{eqnarray}}
\newcommand{\ega}{\end{eqnarray}}
\newcommand{\bb}{\begin{equation}}
\newcommand{\ee}{\end{equation}}
\newtheorem{defn} {Definition}
\newtheorem{te}{Theorem}
\newtheorem{lema}{Lemma}

\usepackage[top=1in, bottom=1in, left=0.75in, right=0.75in]{geometry}
\usepackage{subfigure}
\usepackage[ruled]{algorithm2e}
\usepackage{amsmath}%
\usepackage{amsfonts}%
\usepackage{amssymb}%
\usepackage{graphicx}
\usepackage{url}
\newenvironment{proof}[1][Proof]{\textbf{#1.} }{\ \rule{0.5em}{0.5em}}

\newcommand{\VC}[3]{\ensuremath{#1^{#2}_{#3}}}
\newcommand{\mf}[3]{\ensuremath{\omega _{#3}(#1, #2)}}
\newcommand{\mb}[3]{\ensuremath{\overline{\omega} _{#3}(#1, #2)}}

\newcommand{\etal}{\textit{et al. }}

\begin{document}
\title{Low-Density Parity-Check Codes Which Can Correct Three Errors Under Iterative Decoding}
\author{Shashi~Kiran~Chilappagari,~\IEEEmembership{Student~Member,~IEEE,}~Anantha~Raman~Krishnan,~Bane~Vasic,~\IEEEmembership{Senior Member,~IEEE,}~and~Michael~W.~Marcellin,~\IEEEmembership{Fellow,~IEEE}% <-this % stops a space
\thanks{Manuscript received \today. This work is funded by NSF under Grants CCF-0634969, ECCS-0725405, ITR-0325979 and by the INSIC-EHDR program.}
\thanks{Part of the work was presented at the Information Theory Workshop (ITW), May 5-9, 2008, Porto, Portugal.}
\thanks{S. K. Chilappagari, A. Krishnan, B. Vasic and M. W. Marcellin are with the Department of Electrical and Computer Engineering, University of Arizona, Tucson, Arizona, 85721 USA. (emails: \{shashic, ananthak, vasic, marcellin\}@ece.arizona.edu.}
}% <-this % stops a space
\markboth{Submitted to  IEEE Transactions on Information Theory, October 2008}%
{Submitted to IEEE Transactions on Information Theory, October 2008}
\maketitle
\vspace{-0.4in}
\begin{abstract}
In this paper, we give necessary and sufficient conditions for low-density parity-check (LDPC) codes with column-weight three to correct three errors when decoded using hard-decision message-passing decoding. Additionally, we give necessary and sufficient conditions for column-weight-four codes to correct three errors in four iterations of hard-decision message-passing decoding. We then give a construction technique which results in codes satisfying these conditions. We also provide numerical assessment of code performance via simulation results.
\end{abstract}
\section{Introduction}
First introduced by Gallager \cite{gallager}, LDPC codes have been the focus of intense research in the past decade and many of their properties are now well-understood. The iterative decoding algorithms for LDPC codes have been analyzed in detail, and asymptotic performance results have been derived \cite{richardsonurbanke}. However, estimation of frame-error-rate (FER) for iterative decoding of finite-length LDPC codes is still an unsolved problem. A special case of interest is the performance of iterative decoding at high signal-to-noise ratio (SNR). At high SNRs, a sudden degradation in the performance of iterative decoders has been observed \cite{weaknessmackay},\cite{rich}. This abrupt change manifested in the FER curve is termed as an ``error-floor.''

The error-floor problem is well-understood for iterative decoding over the binary erasure channel (BEC) \cite{di}. Combinatorial structures called ``stopping sets'' were used to characterize the FER for iterative decoding of LDPC codes over the BEC. It was established that decoding failure occurs whenever all the variables belonging to stopping sets are erased. Tian \etal \cite{emd} used this fact to construct irregular LDPC codes which avoid small stopping sets thus improving the guaranteed erasure recovery capability of codes under iterative decoding, and hence improving the error-floors. As in the case of BEC, a strong connection has been found between the existence of low-weight uncorrectable error patterns and error-floors for additive white Gaussian noise (AWGN) channels and binary symmetric channels (BSC) (see \cite{rich} and \cite{chilappagarione}). Hence, studying the guaranteed error correction capability of codes under iterative decoding is important in the context of characterization and improvement of the performance of iterative decoding strategies.

In the past, guaranteed error correction has been approached from the perspective of the decoding algorithm as well as from the perspective of code construction.  Sipser and Spielman \cite{SpielmanExpander} used expansion arguments to derive sufficient conditions for the parallel bit-flipping algorithm to correct a fraction of errors in codes with column-weight greater than four. Burshtein \cite{burshteinBitFlipping} proved that for large enough lengths, almost all codes with column-weights greater than or equal to four can correct a certain fraction of errors under the bit-flipping algorithm. Burshtein and Miller \cite{burshtein} derived the sufficient conditions for message-passing decoding to correct a fraction of errors for codes of column-weight greater than five. However, these proofs were not constructive, i.e., no explicit code construction which satisfied the sufficient conditions was provided. Moreover, the code-lengths required to guarantee the correction of a small number of errors (say $3$) is very high . Also, these arguments cannot be extended for message-passing decoding of codes with column-weight three or four.

In order to construct codes with good error correcting properties under iterative decoding, progressive edge growth (PEG) \cite{peg} and constructions based on finite geometries \cite{LinFiniteGeometry} have been used. However, codes constructed from finite geometries typically have very high column-weight. Although, it has been proved that minimum distance grows at least linearly for codes constructed using PEG, no results proving guaranteed error correction under iterative decoding exist for these codes.

In this work, we derive necessary and sufficient conditions for the correction of three errors in a column-weight-three code under the hard-decision message-passing algorithm. We provide a modified PEG construction which yields codes with such an error-correction capability. Also, we derive the necessary and sufficient conditions for the correction of three errors in four iterations for the case of codes with column-weight four. Again, we provide a modified PEG construction which yields codes with such error-correction capability.

The remainder of the paper is organized as follows: We establish the preliminaries of the work in Section \ref{section2}. The necessary and sufficient conditions for the correction of three errors in column-weight-three codes are derived in Section \ref{section3}. The case of column-weight-four codes is dealt with in Section \ref{section4}. In Section \ref{section5}, we describe a technique to construct codes satisfying the conditions of the theorems and provide numerical results. We conclude with a few remarks in Section \ref{section6}.

\section{Preliminaries}\label{section2}

In this section, we first describe the Tanner graph representation of LDPC codes. Then, we establish the notation that will be used throughout this paper. Finally, we describe the hard-decision message-passing algorithm that will be used for decoding.
\subsection{Notation}
The Tanner graph of an LDPC code, $\mathcal{G}(V, C)$, is a bipartite graph with two sets of nodes: $V$, the variable (bit) nodes and $C$, the check (constraint) nodes. Every edge $e$ in the bipartite graph is associated with a variable node $v$ and a check node $c$. The check nodes (variable nodes, respectively) connected to a variable node (check node, respectively) are referred to as its neighbors.  The degree of a node is the number of its neighbors. In a $(\gamma,\rho)$-regular LDPC code, each variable node has degree $\gamma$ and each check node has degree $\rho$. The girth $g$ is the length of the shortest cycle in $\cal{G}$. Let $S \subset V$ such that $\left | S \right | = y$. If for all choices of $S$, there are at least $z$ neighbors of $S$ in $C$, then we say that the $y \rightarrow z$ condition is satisfied. In this paper, $\bullet$ represents a variable node, $\square$ represents an even-degree check node and $\blacksquare$ represents an odd-degree check node.
\subsection{Hard-Decision Decoding Algorithm}
Let $\mathbf{r} = \left [ r(1), r(2), \dots, r(n) \right ]$, a binary $n$-tuple, be the input to the message-passing decoder. Let $v \in V$ be a variable node with $r(v)$ as its corresponding bit and $c \in C$ be a check node neighboring $v$. Let \mf vcj denote the message that $v$ sends to $c$ in the first half of the $j^{th}$ iteration and \mb cvj denote the message that $c$ sends to $v$ in the second half of the $j^{th}$ iteration

Additionally, let \mf v:j be the set of all messages from a variable $v$ to all its neighboring checks in the first half of the $j^{th}$ iteration. Let \mf v{:\backslash c}j be the set of all messages that a variable node $v$ sends to all its neighboring checks except $c$ in the first half of the $j^{th}$ iteration. Let \mb {:}{v}{j} be the set of all messages received by $v$ from all its neighboring in the second half of the $j^{th}$ iteration. Let \mb {:\backslash c}{v}{j} be the set of all messages received by $v$ from all its neighboring check nodes except $c$ in the second half of the $j^{th}$ iteration. \mb c:j, \mb c{:\backslash v}j, \mf {:}{c}{j} and \mf {:\backslash v}{c}{j} are defined similarly.

The Gallager algorithms \cite{gallager} can be defined as follows: The forward messages, \mf vcj (from variables to checks), are defined as
\begin{equation}\label{Equation:Forward}
\mf vcj = \left \{
\begin{array}{rl}
r(v), & \text{if } j = 1 \\
m, & \text{if } \left| \{ c^\prime : c^\prime \neq c, \mb {c^\prime}{v}{j-1} = m\} \right| \geq b_{v, j} \\
r(v), & \text{otherwise}
\end{array}
\right.
\end{equation}
where $\left| \{ c^\prime : c^\prime \neq c, \mb {c^\prime}{v}{j-1} = m\} \right|$ refers to the total number of messages which are of the value $m \in \{0,1\}$. The backward messages, \mb cvj (from checks to variables), are defined as
\begin{equation}\label{Equation:Backward}
\mb cvj =  \left ( \sum_{m_j \in \mf c{:\backslash v}j } m_j \right ) \text{mod } 2
\end{equation}

At the end of each iteration, an estimate of each variable node is made based on the incoming messages and possibly the received value. The decoder is run until a valid codeword is found or until a maximum number of iterations, say $D$, is reached, whichever is earlier.

In Eqn. (\ref{Equation:Forward}), $b_{v, j}$ is a threshold which is generally a function of the iteration number, $j$, and the degree of the variable $v$. In this paper, we use $b_{v, j} = 2$ for all $v$ and $j$ for decoding column-weight-three codes. For column-weight-four codes, we use $b_{v, j} = 3$ for all $v$ when $1 \leq j \leq 3$ and $b_{v, j} = 2$ for all $v$ when $j \geq 4$.

\textit{Remark:}
We note that Eqns. \ref{Equation:Forward} and \ref{Equation:Backward} then correspond to the Gallager-B algorithm \cite{gallager}. For the Gallager-A algorithm \cite{gallager}, $b_{v, j} = \gamma_v - 1$, for all $j$, where $\gamma_v$ is degree of variable node $v$.\\

\textit{A Note on the Decision Rule:} Different rules to estimate a variable node after each iteration are available, and it is likely that changing the rule after certain number of iterations may be beneficial. However, the analysis of such scenarios is beyond the scope of this paper. Throughout the paper, we use the following decision rule: if all incoming messages to a variable node from neighboring checks are equal, set the variable node to that value; else set it to its received value. 

\subsection{Trapping Sets of the Hard-Decision Decoder}

We discuss briefly the concept of trapping sets. Consider an LDPC code of length $n$. Let $\mathbf{r}$ be the binary vector which is the input to the hard-decision decoder. For output symmetric channels, without loss of generality, we can assume that the all-zero-codeword is transmitted. We make this assumption throughout this paper. The support of a vector $\mathbf{r}$ denoted by $\mathcal{S}(\mathbf{r})$ is defined as the set of all positions $i$ where $r(i) \neq 0$. For each $l$, $1 \leq l \leq D$, let $\mathbf{x}^l$ be the codeword estimate of the decoder at the end of the $l^{th}$ iteration.  A variable node $v$ is said to be \textit{eventually correct} if there exists a positive integer $l_c$ such that for all $l \geq l_c$, $v$ does not belong to $\mathcal{S}(\mathbf{x}^l)$.
\begin{defn}\cite{rich}
A decoding failure is said to have occurred if there does not exist $l \leq D$ such that $\mathcal{S}(\mathbf{x}^l) = \emptyset$
\end{defn}
\begin{defn}\cite{rich}
Let $\mathbf{T}(\mathbf{r})$ denote the set of variable nodes that are not eventually correct. If $\mathbf{T}(\mathbf{r})$ is not empty, let $a= \left | \mathbf{T}(\mathbf{r}) \right|$ and $b$ be the number of odd-degree check nodes in the subgraph induced by $\mathbf{T}(\mathbf{r})$. We say that $\mathbf{T}(\mathbf{r})$ is an $(a,b)$ trapping set.
\end{defn}
\begin{defn}
Let $\mathcal{T}$ be a trapping set and let $\mathbf{R}(\mathcal{T}) = \{ \mathbf{r}: \mathbf{T}(\mathbf{r}) = \mathcal{T} \}$. The critical number $\xi$ of trapping set $\mathcal{T}$ is the minimum number of variable nodes that have to be initially in error for the decoder to end up in the trapping set $\mathcal{T}$. That is, $\xi = \min_{\mathbf{r} \in \mathbf{R}(\mathcal{T})} \left | \mathcal{S}(\mathbf{r}) \right |$.
\end{defn}
\begin{defn}\cite{rich}
Let $\mathcal{T}$ be a trapping set. If $\mathbf{T}(\mathbf{r}) = \mathcal{T}$, then $\mathcal{S}(\mathbf{r})$ is a failure set of $\mathcal{T}$.
\end{defn}
\begin{defn}
For transmission over the BSC, $\mathbf{r}$ is a fixed point of the decoding algorithm if $\mathcal{S}(\mathbf{r}) = \mathcal{S}(\mathbf{x}^l)$ for all $l$.
\end{defn}
It follows that for transmission over the BSC, if $\mathbf{r}$ is a fixed point, then $\mathbf{T}(\mathbf{r}) = \mathcal{S}(\mathbf{r})$ is a trapping set. Now, we have the following theorem which provides the sufficient condition for a set of variables to be a trapping set:
\begin{te}\label{thm1}\cite{colwtthreepaper}
Let $\mathcal{G}(V, C)$ be the Tanner graph of a column-weight-three code. Let $\mathcal{T} \subset V$, be a set consisting of $v$ variable nodes with induced subgraph $\cal{I}$. Let the checks in $\cal{I}$ be partitioned into two disjoint subsets, namely, $\cal{O}$ consisting of checks with odd degree and $\cal{E}$ consisting of checks with even degree. If (a) every variable node in $\cal{I}$ is connected to at least two  checks in $\cal{E}$ and at most one check in $\cal{O}$ and (b) no two checks of $\cal{O}$ are connected to the same variable node outside $\cal{I}$, then $\mathcal{T}$ is a trapping set.
\end{te}
\begin{proof}
See \cite{colwtthreepaper}.
\end{proof}

\section{Column-Weight-Three Codes} \label{section3}
In this section, we establish necessary and sufficient conditions for a column-weight-three code to correct three errors. We first illustrate three trapping sets and show that the critical number of these trapping sets is three thereby providing necessary conditions to correct three errors. We then prove that avoiding structures isomorphic to these trapping sets in the Tanner graph is sufficient to guarantee correction of three errors.

Fig. \ref{trappingsets} shows three subgraphs induced by different numbers of variable nodes. Let us assume that in all these induced graphs, no two odd degree checks are connected to the same variable node outside the graph. By the conditions of Theorem \ref{thm1}, all these induced subgraphs are trapping sets. Fig. \ref{sixcycle} is a $(3,3)$ trapping set, Fig. \ref{53trappingset} is a $(5,3)$ trapping set and Fig. \ref{weight8codeword} is a $(8,0)$ trapping set. Note that a $(3,3)$ trapping set is isomorphic to a six-cycle and the $(8,0)$ trapping set is a codeword of weight eight. We now have the following result:
\begin{figure*}[htb]
\centering
\subfigure[] % caption for subfigure a
{\label{sixcycle}\includegraphics[width=0.18\textwidth]{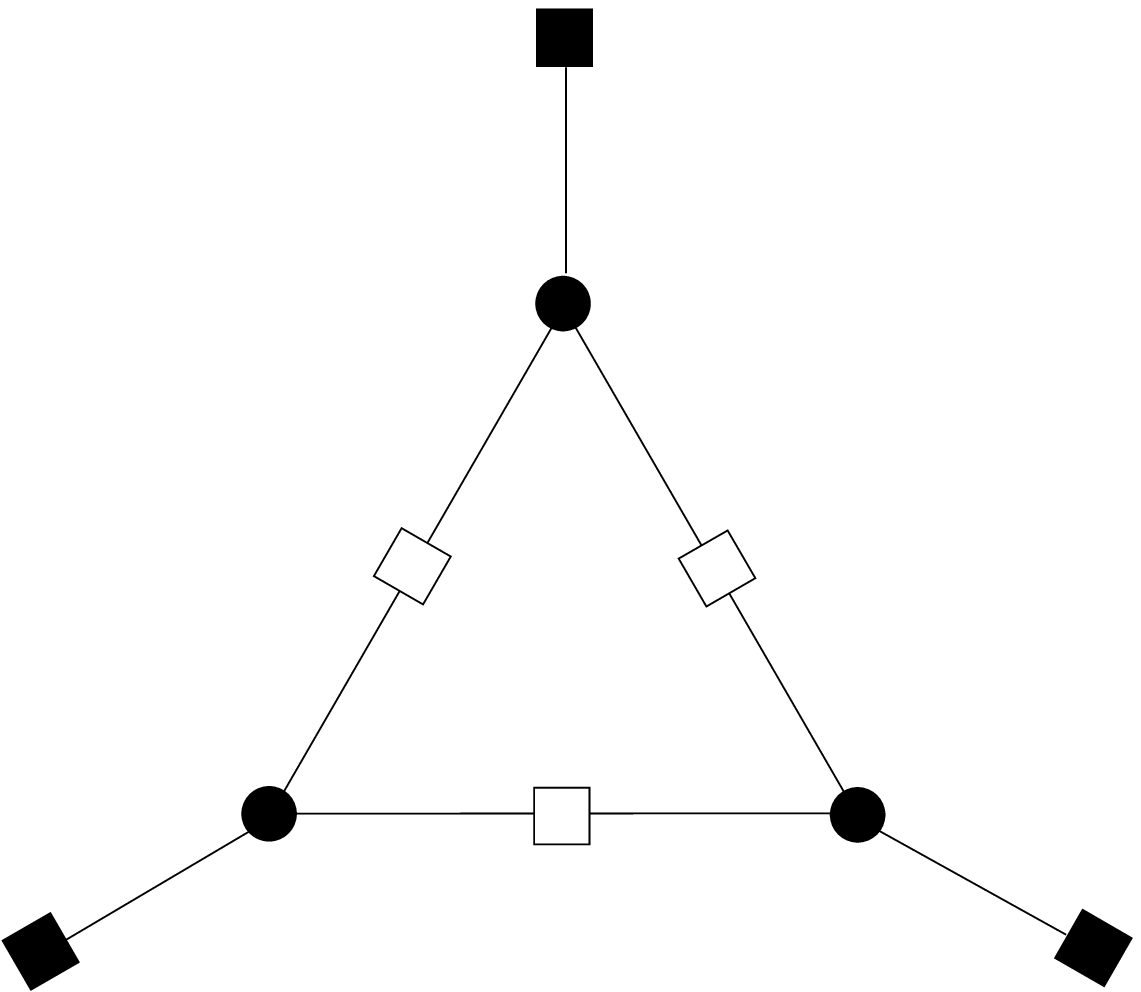} }
\hspace{0.05\textwidth}
\subfigure[] % caption for subfigure a
{\label{53trappingset}\includegraphics[width=0.3\textwidth]{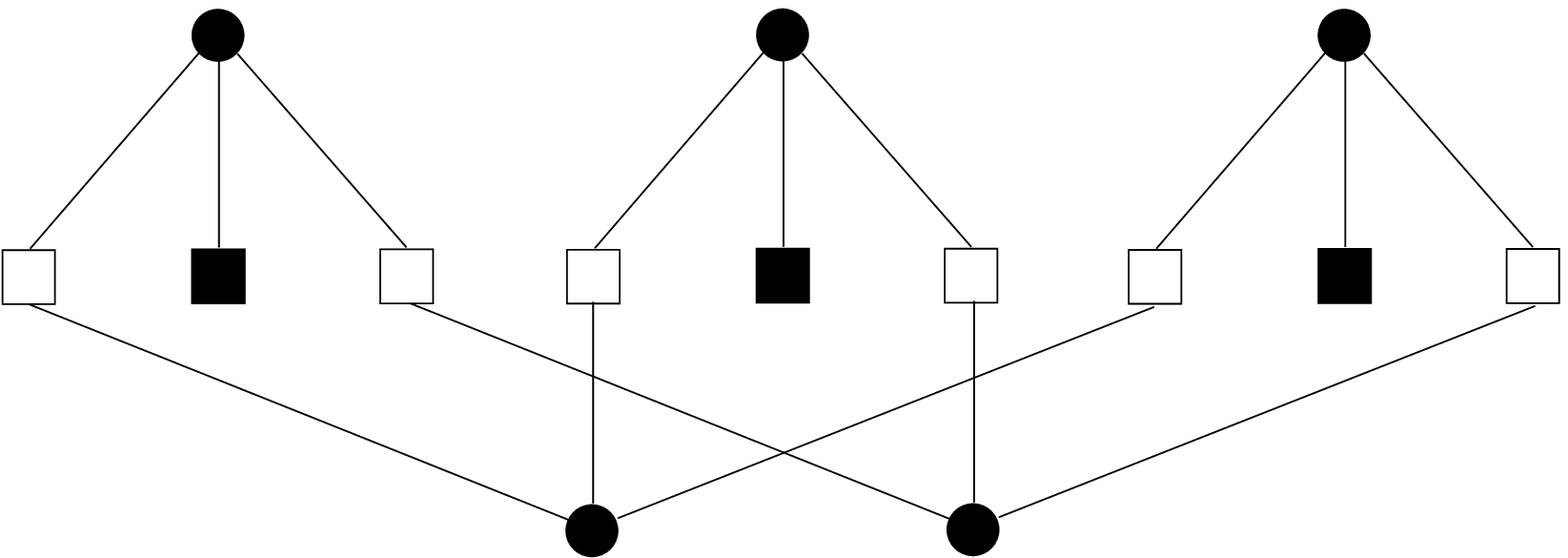} }
\hspace{0.05\textwidth}
\subfigure[]  % caption for subfigure b
{\label{weight8codeword}\includegraphics[width=0.25\textwidth]{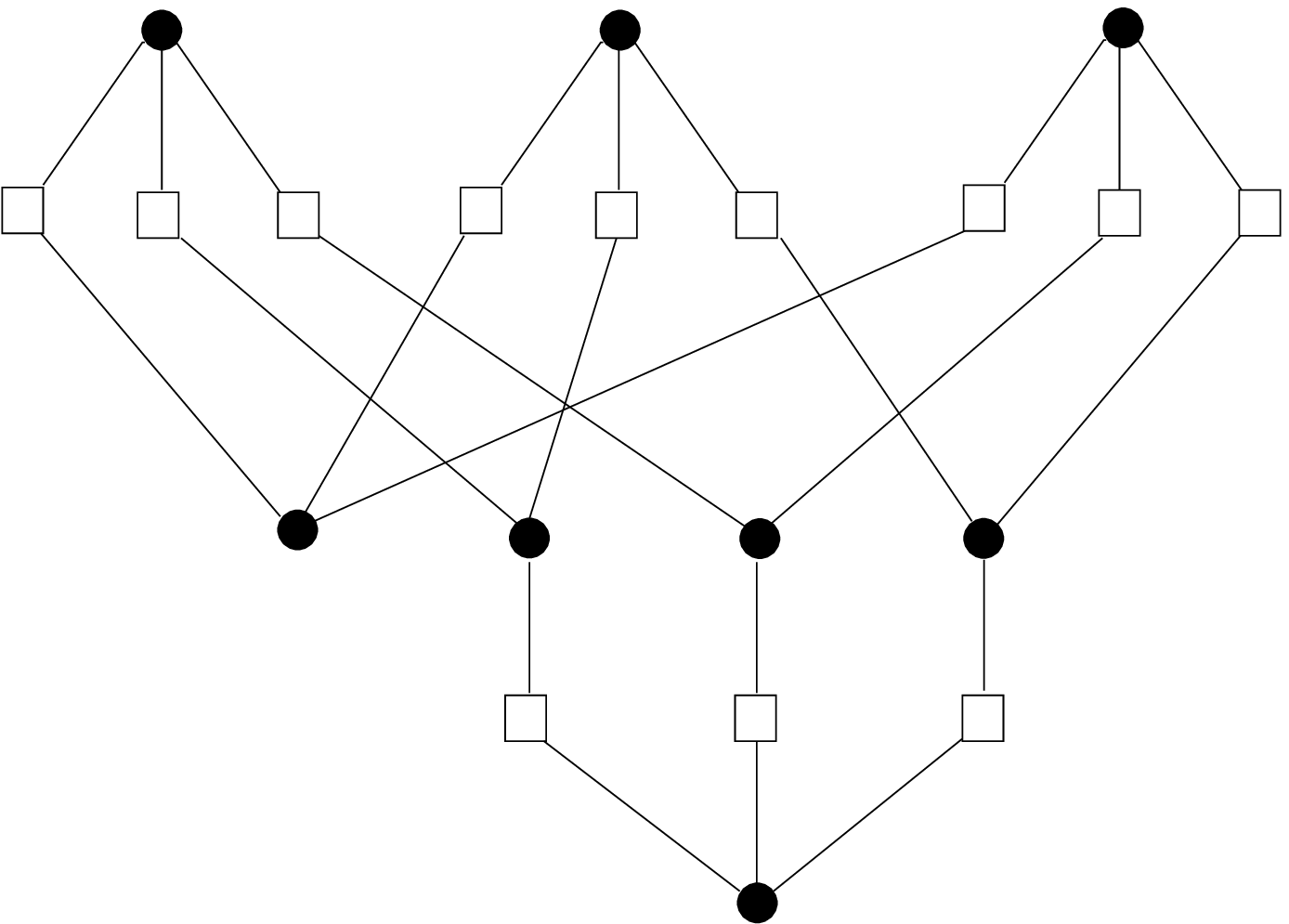} }
\caption{Examples of trapping sets with critical number three: \protect\subref{sixcycle} a $(3,3)$ trapping set; \protect\subref{53trappingset} a $(5,3)$ trapping set; and \protect\subref{weight8codeword} an $(8,0)$ trapping set.}\label{trappingsets}
\end{figure*}

\begin{lema}
The critical number for a $(3,3)$ trapping set which is also a fixed point is at most three. There exist (5,3) and (8,0) trapping sets with critical number three and no (5,3) or (8,0) trapping sets with critical number less than three.  
\end{lema}
\begin{proof}
\begin{figure*}[htb]
\centering
\subfigure[] % caption for subfigure a
{
    \label{vc1}
\includegraphics[width=0.43\textwidth]{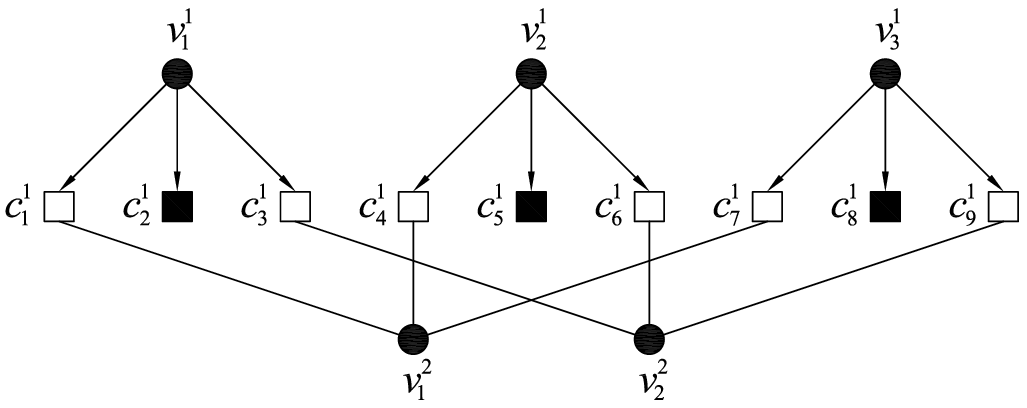}
}
\hspace{0.07\textwidth}
\subfigure[] % caption for subfigure a
{
    \label{cv1}
\includegraphics[width=0.43\textwidth]{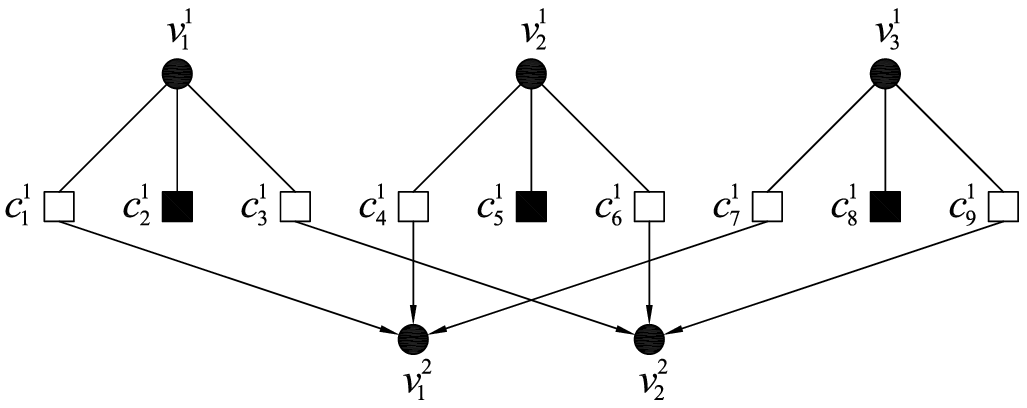}
}
\subfigure[]  % caption for subfigure b
{
    \label{vc2}
\includegraphics[width=0.43\textwidth]{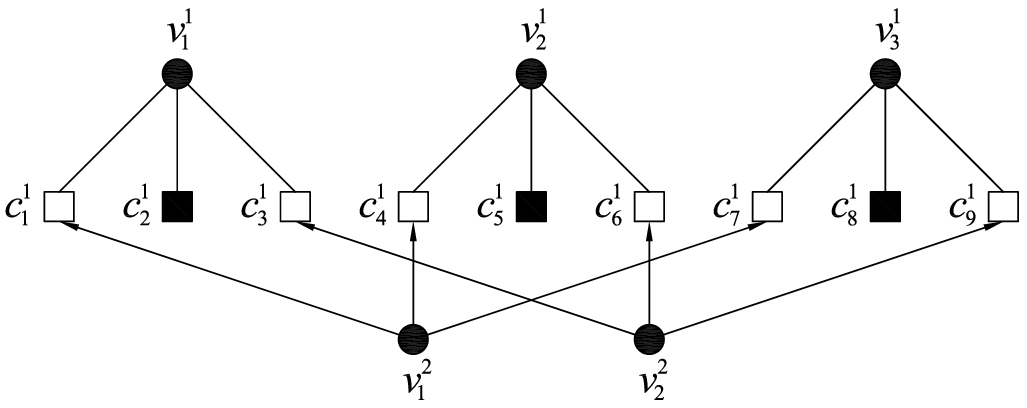}
}
\hspace{0.07\textwidth}
\subfigure[]  % caption for subfigure b
{
    \label{cv2}
\includegraphics[width=0.43\textwidth]{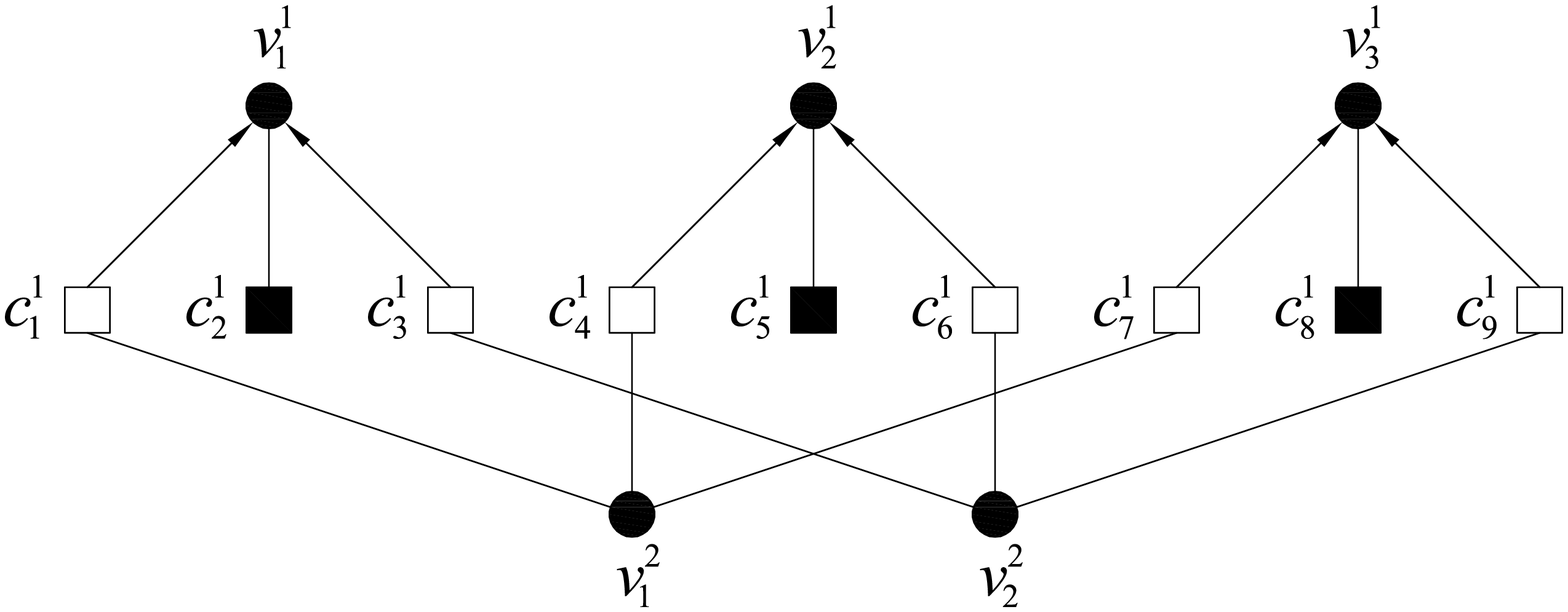}
}
 \caption{Illustration of message passing for a $(5,3)$ trapping set: \subref{vc1} variable to check messages in round one; \subref{cv1} check to variable messages in round one; \subref{vc2} variable to check messages in round two; and \subref{cv2} check to variable messages in round two. Arrow-heads indicate the messages with value 1.}\label{messagepassing}
\end{figure*}
The proof for $(3,3)$ case is trivial. We prove the lemma for the case of $(5,3)$ trapping sets and omit the proof for $(8,0)$ trapping sets.

Consider the $(5, 3)$ trapping set shown in Fig. \ref{messagepassing}. Let $V^1 := \left \{ \VC v11, \VC v12, \VC v13 \right \}$ be the set of variables which are initially in error. Let $C^1 := \{\VC c11, \VC c12, \dots, \VC c19\}$ and $V^2 := \{ \VC v21, \VC v22\}$. Also, assume that no variable node in $V \backslash (V^1 \cup V^2)$, has two or more neighbors in $C^1$. In the first iteration, we have:
\begin{eqnarray}
\mf vc1 & = & \left \{
\begin{array}{rl}
1 & \text{if }v \in V^1 \\
0 & \text{otherwise}
\end{array}
\right. \label{Eqn:53mf1}\\
\mb cv1 & = & \left \{
\begin{array}{rl}
1 & \text{if }c \in C^1,~v \notin V^1 \\
0 & \text{otherwise}
\end{array}
\right.\label{Eqn:53mb1}
\end{eqnarray}
Consequently, all variable nodes in $V^2$ are decoded incorrectly at the end of the first iteration. In the second iteration:
\begin{eqnarray}
\mf vc2 & = & \left \{
\begin{array}{rl}
1 & \text{if }v \in V^2 \\
0 & \text{otherwise}
\end{array}
\right. \\
\mb cv2 & = & \left \{
\begin{array}{rl}
1 & \text{if }c \in C^1 \backslash \{\VC c12, \VC c15, \VC c18\},~v \notin V^2 \\
0 & \text{otherwise}
\end{array}
\right.
\end{eqnarray}
and all variable nodes in $V^1$ are decoded incorrectly. Continuing in this fashion, $\mf vc3 = \mf vc1$ and $\mb cv3 = \mb cv1$. That is, the messages being passed in the Tanner graph would repeat after every two iterations. Hence, three variable nodes in error initially can lead to a decoder failure and therefore, this $(5, 3)$ trapping set has critical number equal to three.
\end{proof}
\begin{te}
To guarantee that three errors in a column-weight-three LDPC code can be corrected by the Gallager-A algorithm, it is necessary to avoid $(3,3)$, $(5,3)$ and $(8,0)$ trapping sets in its Tanner graph.
\end{te}
\begin{proof}
Follows from the discussion above.
\end{proof}
We now state and prove the main theorem.
\begin{te}
If the Tanner graph of a column-weight-three LDPC codes has girth eight and does not contain a subgraph isomorphic to a $(5,3)$ trapping set or a subgraph isomorphic to an $(8,0)$ trapping set, then any three errors can be corrected using the Gallager-A algorithm.
\end{te}
\begin{proof}
Let $V^1 := \{\VC v11, \VC v12, \VC v13\}$ be the three erroneous variables and $C^1$ be the set of the checks connected to the variables in $V^1$. In a column-weight-three code (free of cycles of length four) the variables in $V^1$ can induce only one of the five subgraphs given in Fig. \ref{errorconfigs}. In each case, $\mf v:1 = \{1\}$ if $v \in V^1$ and is $0$ otherwise. The proof proceeds by examining these subgraphs one at a time and proving the correction of the three erroneous variables in each case.
\begin{figure*}[htb]
\centering
\subfigure[] % caption for subfigure a
{
    \label{config1}
\includegraphics[width=0.30\textwidth]{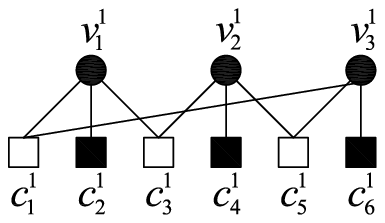}
}
%\hspace{0.05\textwidth}
\subfigure[] % caption for subfigure a
{
    \label{config2}
\includegraphics[width=0.30\textwidth]{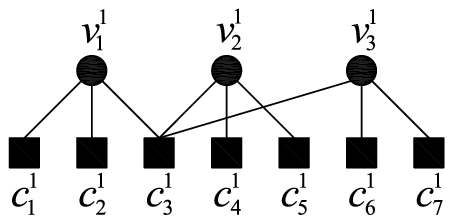}
}
\subfigure[]  % caption for subfigure b
{
    \label{config3}
\includegraphics[width=0.30\textwidth]{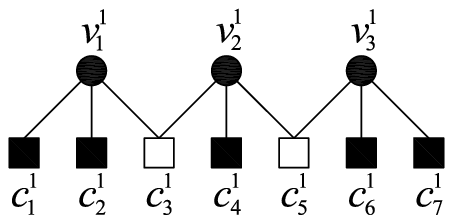}
}
\subfigure[]  % caption for subfigure b
{
    \label{config4}
\includegraphics[width=0.30\textwidth]{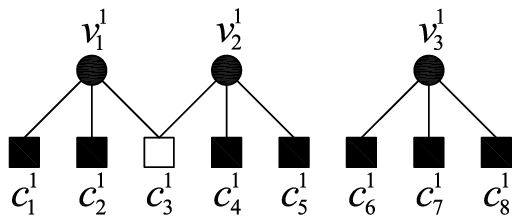}
}
\subfigure[]  % caption for subfigure b
{
    \label{config5}
\includegraphics[width=0.30\textwidth]{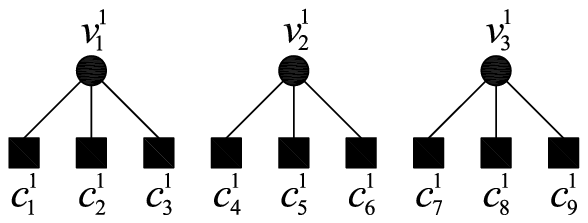}
}
 \caption{All the possible subgraphs that can be induced by three variable nodes in a column-weight-three code}\label{errorconfigs}
\end{figure*}
 
\textbf{Subgraph 1:} Since the girth of the code is eight, it has no six cycles. Hence, the configuration in Fig. \ref{config1} is not possible. 

\textbf{Subgraph 2:} The variables in $V^1$ induce the subgraph shown in Fig. \ref{config2}. At the end of the first iteration: 
\begin{equation}
\mb cv1 = \left \{
\begin{array}{rl}
1 & \text{if }c \in C^1,~v \notin V^1 \\
0 & \text{otherwise}
\end{array}
\right.
\end{equation}
There cannot exist a variable node which is connected to two or more checks in the set $C^1$ without introducing either a six-cycle or a subgraph isomorphic to $(5,3)$ trapping set. At the end of first iteration, $\mb :v1 = \{0\}$ for all $v \in V^1$. Furthermore, there exists no $v \notin V^1$ for which $\mb :v1 = \{1\}$. Hence, if a decision is made after the first iteration, a valid codeword is found and the decoder is successful.

\textbf{Subgraph 3:} The variables in $V^1$ induce the subgraph shown in Fig. \ref{config3}. At the end of the first iteration:
\begin{equation}
\mb cv1 = \left \{
\begin{array}{rl}
1 & \text{if } c \in C^1 \backslash \{ \VC c13, \VC c15 \},~v \notin V^1 \\
1 & \text{if }c \in \{\VC c13, \VC c15 \},~v \in V^1 \\
0 & \text{otherwise}
\end{array}
\right.
\end{equation}
For no $v \notin V^1$, $\mb:v1 = \{1\}$ as this would introduce a four-cycle or a six-cycle in the graph. For any $v \notin V^1$, $\mf vc2 = 1$ only if $\mb {:\backslash c}v1 = \{1\}$. This implies that $v$ has two checks in $C^1 \backslash \{ \VC c13, \VC c15 \}$. Let $V^2$ be the set of such variables. We have the following lemma:
\begin{lema}
There can be at most one variable in $V^2$.
\end{lema}
\begin{proof}
Suppose $\left | V^2 \right | = 2$. Specifically, assume $V^2 = \{ \VC v21, \VC v22\}$. The proof is similar for $\left | V^2 \right | > 2$. First note that for any $v \in V^2$, $v$ cannot be connected to \VC c14 as it would create a six-cycle. Next, let $\VC C11 := \{ \VC c11, \VC c12\}$ and $\VC C12 := \{ \VC c16, \VC c17\}$. Then, $v$ cannot have both checks in either \VC C11 or \VC C12 as this would cause a four-cycle. Hence, $v$ has one check in \VC C11 and one check in \VC C12. Assume without loss of generality that \VC v21 is connected to \VC c11 and \VC c16. Then, \VC v22 cannot be connected to \VC c11 and \VC c17 as this would form a six-cycle. \VC v22 cannot be connected to \VC c12 and \VC c17 as it would create a $(5, 3)$ trapping set. Hence, $\left | V^2 \right | < 2$.
\end{proof}

Let $\VC v21 \in V^2$ be connected to \VC c11, \VC c16 and an additional check \VC c21. In the second iteration:
\begin{eqnarray}
\mf vc2 & = & \left \{
\begin{array}{rl}
1 & \text{if } v \in \{ \VC v11, \VC v13\},~c \notin \{\VC c13, \VC c15 \} \\
1 & \text{if } v = \VC v12 \\
1 & \text{if } v = \VC v21,~c = \VC c21 \\
0 & \text{otherwise}
\end{array}
\right. \\
\mb cv2 & = & \left \{
\begin{array}{rl}
1 & \text{if } c \in C^1 \backslash \{ \VC c14 \},~v \notin V^1 \\
1 & \text{if } c \in \{ \VC c13, \VC c14, \VC c15 \},~v \neq \VC v12 \\
1 & \text{if } c = \VC c21,~v \neq \VC v21 \\
0 & \text{otherwise}
\end{array}
\right.
\end{eqnarray}
We have the following lemma:
\begin{lema}\label{Lemma3:TwoIncorrectMessages}
There cannot exist any variable $v \notin V^1 \bigcup V^2$ such that it receives two or more incorrect messages at the end of the second iteration.
\end{lema}
\begin{proof}
Suppose there existed a variable $v$ such that it received two incorrect messages in the second iteration. Then, it would be connected to two checks in the set $C^1 \bigcup \{ \VC c21 \}$. This is not possible as it would introduce a four-cycle, six-cycle or a $(5, 3)$ trapping set (\textit{e.g.} if $v$ is connected to \VC c14 and \VC c21, it would form a $(5, 3)$ trapping set).
\end{proof}

Thus, in the third iteration:
\begin{eqnarray}
\mf vc3 & = & \left \{
\begin{array}{rl}
1 & \text{if } v \in \{ \VC v11, \VC v13\},~c \notin \{ \VC c13, \VC c15\} \\
1 & \text{if } v = \VC v21,~c = \VC c21 \\
0 & \text{otherwise}
\end{array}
\right. \\
\mb cv3 & = & \left \{
\begin{array}{rl}
1 & \text{if } c \in \{ \VC c11, \VC c12, \VC c16, \VC c17 \},~v \notin \{ \VC v11, \VC v13\} \\
1 & \text{if } c = \VC c21,~v \neq \VC v21 \\
0 & \text{otherwise}
\end{array}
\right.
\end{eqnarray}
At the end of the third iteration, $\mb :v3 = \{0\}$ for all $v \in V^1$. Also, we have the following lemma:
\begin{lema}
There exists no $v \notin V^1$ such that $\mb :v3 = \{1\}$.
\end{lema}
\begin{proof}
Suppose there exists $v$ such that $\mb :v3 = \{1\}$. Then, $v$ is connected to three checks in the set $\{\VC c11, \VC c12, \VC c16, \VC c17, \VC c21 \}$. This implies that $\mb :v2 = \{1\}$. However, from Lemma \ref{Lemma3:TwoIncorrectMessages} it is evident that no such $v$ exists.
\end{proof}
Hence, if a decision is made after the third iteration, a valid codeword is found and the decoder is successful.

\textbf{Subgraph 4:}  The variables in $V^1$ induce the subgraph shown in Fig. \ref{config4}. At the end of the first iteration:
\begin{equation}
\mb cv1 = \left \{
\begin{array}{rl}
1 & \text{if } c \in C^1 \backslash \{ \VC c13 \},~v \notin V^1 \\
1 & \text{if } c = \VC c13,~v \in \{ \VC v11,\VC v12 \} \\
0 & \text{otherwise}
\end{array}
\right.
\end{equation}
For no $v \in V \backslash V^1$, $\mb :v2 = \{1\}$. For any $v \in V \backslash V^1$, $\mf vc2 = 1$ only if $\mb {:\backslash c}v1 = \{1\}$. Let $V^2$ be the set of all such variables. We have the following lemma:
\begin{lema}
(i) $V^2$ has at most four variables, and (ii) No two variables in $V^2$ can share a check in $C \backslash C^1$.
\end{lema}

\textit{Sketch of the Proof}: There exists no variable which is connected to two checks from the set $\{ \VC c11, \VC c12, \VC c13, \VC c14, \VC c15 \}$ as it would introduce a four-cycle or a six-cycle. However, a variable node can be connected to one check from $\{\VC c11, \VC c12, \VC c13, \VC c14, \VC c15\}$ and to one check from $\{\VC c16, \VC c17, \VC c18 \}$. There can be at most four such variable nodes. When four such variable nodes exist, none are connected to \VC c13. Also, these four variable nodes cannot share checks outside the set $C^1\backslash \{ \VC c13 \} $. 

Let these four variable nodes be labeled \VC v21, \VC v22, \VC v23 and \VC v24 and their third checks \VC c21, \VC c22, \VC c23 and \VC c24, respectively. Let $C^2 := \{ \VC c21, \VC c22, \VC c23, \VC c24 \}$. Hence, in the second iteration:
\begin{eqnarray}
\mf vc2 & = & \left \{
\begin{array}{rl}
1 & \text{if } v \in \{ \VC v11, \VC v12\},~c \neq \VC c13 \\
1 & \text{if } v \in V^2,~c \in C^2 \\
0 & \text{otherwise}
\end{array}
\right. \\
\mb cv2 & = & \left \{
\begin{array}{rl}
1 & \text{if } c \in \{\VC c11, \VC c12, \VC c14, \VC c15 \},~v \notin \{\VC v11, \VC v12\}  \\
1 & \text{if } c \in C^2 ,~v \notin V^2 \\
0 & \text{otherwise}
\end{array}
\right.
\end{eqnarray}
At the end of the second iteration $\mb :v2 = \{0\}$ for all $v \in V^1$. Moreover, for no $v \notin V^1$, $\mb :v2 = \{1\}$. So, if a decision is made after the second iteration, a valid codeword is reached and the decoder is successful. 

\textbf{Subgraph 5:} The variables in $V^1$ induce the subgraph shown in Fig. \ref{config5}. At the end of the first iteration:
\begin{equation}
\mb cv1 = \left \{
\begin{array}{rl}
1 & \text{if }c \in C^1 ,~v \notin V^1 \\
0 & \text{otherwise}
\end{array}
\right.
\end{equation}
If there exists no variable $v \in V \backslash V^1$ such that $\mb :v1 = \{1\}$, a valid codeword is reached after the first iteration. Suppose this is not the case. Let $V^2$ be the set of variables which receive two or more incorrect messages. Then, we have the following lemma:
\begin{lema}
(i) There exists one variable $\VC v21 \in V^2$ such that $\mb :{\VC v21}1 = \{1\}$, and (ii) $V^2$ has at most three variables which receive two incorrect messages at the end of the first iteration. Furthermore, they cannot share a check in $C \backslash C^1$.
\end{lema}
\begin{proof}
We omit the proof of Part (ii) as it is straightforward. Part (i) is proved as follows: If there existed no variable, \VC v21, such that $\mb :{\VC v21}1 = \{1\}$, then the decoder would converge in one iteration. Next, suppose $\VC v21, \VC v22 \in V^2$ such that $\mb :{v^2_1}1 = \mb :{v^2_2}1 = \{1\}$. Without loss of generality, let \VC v21 be connected to $\VC c11,\VC c14$ and $\VC c17$. Then, \VC v21 would share two checks in the set $C^1$. It is thennot possible to connect \VC v22 without introducing a six-cycle or a $(5, 3)$ trapping set (\textit{e.g.}, if \VC v21 is connected to \VC c12, \VC c15 and \VC c18, then it would introduce a $(5, 3)$ trapping set).
\end{proof}

Let the third checks connected to \VC v22, \VC v23 and \VC v24 be \VC c21, \VC c22 and \VC c23, respectively and let $C^2 := \{\VC c21, \VC c22, \VC c23\}$
In the second iteration:
\begin{eqnarray}
\mf vc2 & = & \left \{
\begin{array}{rl}
1 & \text{if }v = \VC v21 \\
1 & \text{if }v \in V^2 \backslash \{ \VC v21 \},~c \in C^2 \\
0 & \text{otherwise}
\end{array}
\right. \\
\mb cv2 & = & \left \{
\begin{array}{rl}
1 & \text{if }c \in \{ \VC c11, \VC c13, \VC c17\},~v \neq \VC v21 \\
1 & \text{if }c \in C^2,~v \notin V^2 \\
0 & \text{otherwise}
\end{array}
\right.
\end{eqnarray}
There cannot exist a variable node which is connected to one check from $C^2$ and to one check from $\{\VC v11, \VC v14, \VC v17\}$. Also, there cannot be a variable node which is connected to all three checks in the set $C^2$  as this would introduce a graph isomorphic to the $(8,0)$ trapping set. However, there can be at most two variable nodes which receive two incorrect messages from the checks in $C^2$, say \VC v31 and \VC v32. Let the third checks connected to them be \VC c31 and \VC c32, respectively. Let $V^3 := \{ \VC v31, \VC v32 \}$ and $C^3 := \{\VC c31, \VC c32 \}$. At the end of the second iteration, variables \VC v11, \VC v12 and \VC v13 receive one incorrect message each. Variables in the set $V^3$ receive two incorrect messages each. Therefore, in the third iteration, we have:
\begin{eqnarray}
\mf vc3 & = & \left \{
\begin{array}{rl}
1 & \text{if }v \in V^1,~c \notin \{ \VC c11, \VC c14, \VC c17 \} \\
1 & \text{if }v \in V^3,~c \in C^3 \\
0 & \text{otherwise}
\end{array}
\right. \\
\mb cv3 & = & \left \{
\begin{array}{rl}
1 & \text{if } c \in C^1\backslash \{ \VC c11, \VC c14, \VC c17 \},~v \notin V^1\\
1 & \text{if } c \in C^3,~v \notin V^3
\end{array}
\right.
\end{eqnarray}
At the end of the third iteration, $\mb:v3 = \{1\}$ for all $v \in V^1$. Furthermore, for no $v \notin V^1$, $\mb :v3 = \{1\}$. So, if a decision is made after the third iteration, a valid codeword is reached and the decoder is successful.
\end{proof}

\section{Column-Weight-Four Codes}\label{section4}
In this section, we derive necessary and sufficient conditions for the correction of three errors in column-weight-four codes in four iterations of iterative decoding. This result is inspired by the analysis of error events in high-rate codes with column-weight four. In simulations, it was found that received vectors which did not converge to a valid codeword in the first 4 to 5 iterations did not converge thenceforth. Hence, it is desirable to devise codes and decoding strategies in which vectors having a small number of errors converged rapidly to a codeword. To this end, it was found that a hybrid decoding strategy could correct three errors in four iterations if certain conditions are satisfied by the code. This result is summarized as follows:
\begin{te}\label{Theorem:4:1}
An LDPC code with column-weight four and girth six can correct three errors in four iterations of message-passing decoding if and only if the conditions, $4 \rightarrow 11$, $5 \rightarrow 12$, $6 \rightarrow 14$, $7 \rightarrow 16$ and $8 \rightarrow 18$ are satisfied.
\end{te}

\textit{Remark:} It is worth noting that if a graph of girth six satisfies the $4 \rightarrow 11$ condition, then it satisfies the $5 \rightarrow 12$ condition as well. However, the addition of this extra constraint aids in the proof of the theorem.

\begin{proof}
First, we prove the sufficiency of the conditions of Theorem \ref{Theorem:4:1}.

Let $V^1 := \{\VC v11, \VC v12, \VC v13\}$ be the three erroneous variables. Let $C^1$ be the set of checks that are connected to the variables in $V^1$. The variables in $V^1$ can induce only one of the five subgraphs shown in Fig. \ref{Figure4}. We prove that in each case, the decoding algorithm converges to the correct codeword in four iterations. 

\begin{figure*}[!htb]
\centering
\subfigure[]
{ \label{Figure4a} \includegraphics[width=0.5\textwidth]{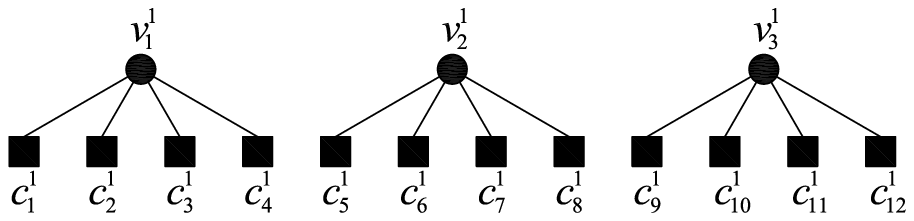} }\\
\subfigure[]{\label{Figure4b}\includegraphics[width=0.49\textwidth]{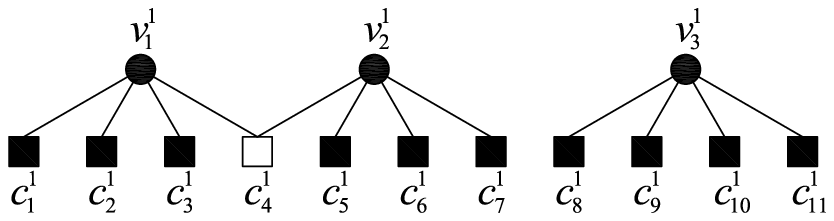}}
\subfigure[]{\label{Figure4c}\includegraphics[width=0.49\textwidth]{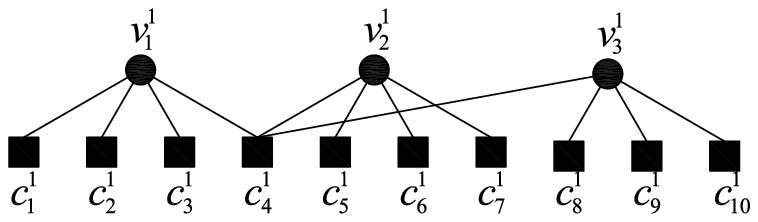} } \\
\subfigure[]{\label{Figure4d}\includegraphics[width=0.49\textwidth]{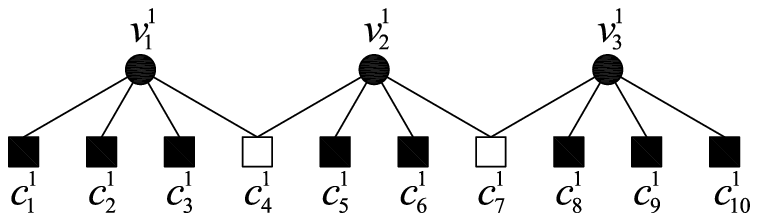}}
\subfigure[]{\label{Figure4e}\includegraphics[width=0.49\textwidth]{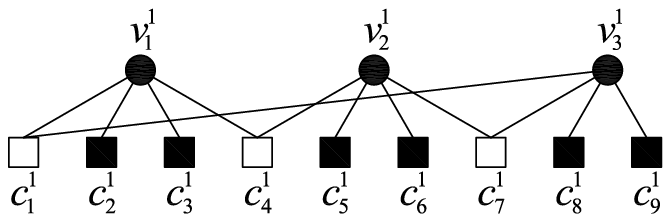}}
\caption{All the possible subgraphs that can be induced by three variable nodes in a column-weight-four code}\label{Figure4}
\end{figure*}

\textbf{Subgraph 1}: The variables in $V^1$ induce the subgraph shown in Fig. \ref{Figure4a}. At the end of the first iteration, $\mb :v1 = \{0\}$ for all $v \in V^1$. Moreover, no variable receives four incorrect messages after the first iteration as the existence of such a variable node would create a four-cycle. If a decision is made after the first iteration, the decoder is successful. 

\textbf{Subgraph 2}: The variables in $V^1$ induce the subgraph shown in Fig. \ref{Figure4b}. At the end of the first iteration:
\begin{equation}
\mb cv1 = \left \{
\begin{array}{rl}
1 & \text{if }c \in C^1 \backslash \{ \VC c14\},~v \notin V^1 \\
1 & \text{if }c = \VC c14,~v \in \{\VC v11, \VC v12 \} \\
0 & \text{otherwise}
\end{array}
\right.
\end{equation}
For no $v \in V \backslash V^1$, $\mb:v1 = \{1\}$ as it would introduce a four-cycle. For any $v \notin V^1$, $\mf vc2 = 1$ only if $\mb{:\backslash c}v1 = \{1\}$. This implies that $v$ is connected to three checks in $C^1 \backslash \{ \VC c14\}$. Let $V^2$ denote the set of such variables. We have the following lemma:
\begin{lema}There can be at most three variables in $V^2$. Furthermore, no two variable nodes in $V^2$ share any check in the set $C \backslash C^1$.
\end{lema}
\begin{proof}
Let $V^2 = \{ \VC v21, \VC v22, \VC v23, \VC v24\}$. Then the set of variable nodes $V^1 \bigcup V^2$ has at most 15 neighboring checks. This violates the $7 \rightarrow 16$ condition. Hence, $V^2$ can have at most three variables. Next, let $\VC v21, \VC v22 \in V^2$. Suppose they share a fourth check $c$. Since \VC v13 can share at most two checks with \VC v21 and \VC v22, assume that \VC c1{10} and \VC c1{11} are not neighbors of \VC v21, \VC v22. The neighbors of the variable nodes in the set $\{ \VC v11, \VC v12, \VC v21, \VC v22\}$ all belong to the set $\{\VC c11,\dots, \VC c19 \} \bigcup \{ \VC c21\}$ which has cardinality 10, thus violating the $4 \rightarrow 11$ condition.
\end{proof}

Let the fourth neighboring checks of \VC v21, \VC v22 and \VC v23 be \VC c21, \VC c22 and \VC c23, respectively. Let $C^2 = \{\VC c21, \VC c22, \VC c23\}$. In the second iteration:
\begin{eqnarray}
\mf vc2 & = & \left \{
\begin{array}{rl}
1 & \text{if }v \in \{\VC v11, \VC v12\},~c \neq \VC c14 \\
1 & \text{if }v \in V^2,~c \in C^2 \\
0 & \text{otherwise}
\end{array}
\right. \\
\mb cv2 & = & \left \{
\begin{array}{rl}
1 & \text{if }c \in \{\VC c11, \VC c12, \VC c13, \VC c15, \VC c16, \VC c17\},~v \in V^1 \\
1 & \text{if }c \in C^2,~v \notin V^2 \\
0 & \text{otherwise}
\end{array}
\right.
\end{eqnarray}
For all $v \in V^1$, $\mb:v2 = \{0\}$. For no $v \in V^2$, $\mb :v2 = \{1\}$. We now have the following lemma:
\begin{lema}
There exists no variable $v \notin V^1 \bigcup V^2$ such that $\mb :v2 = \{1\}$.
\end{lema}
\begin{proof}
The proof is by contradiction. Let $v \notin V^1 \bigcup V^2$ such that $\mb :v2 = \{1\}$. Then, $v$ is connected to four checks in $\{ \VC c11, \VC c12, \VC c13, \VC c15, \VC c16, \VC c17\} \bigcup C^2$. Note that only two neighbors of $v$ can belong to $\{ \VC c11, \VC c12, \VC c13, \VC c15, \VC c16, \VC c17\}$ without introducing a four-cycle. This combined with the fact that there are at most three variable nodes in $V^2$ implies that there are only two cases:\\
(a) $v$ has two neighbors in $\{ \VC c11, \VC c12, \VC c13, \VC c15, \VC c16, \VC c17\}$ and two neighbors in $C^2$, say \VC c21 and \VC c22. In this case, the set of variable nodes $V^1 \bigcup \{\VC v21, \VC v22, v\}$ has $13$ check nodes, violating the $6 \rightarrow 14$ condition.

(b) $v$ has one neighbor in $\{ \VC c11, \VC c12, \VC c13, \VC c15, \VC c16, \VC c17\}$ and three neighbors in $C^2$. In this case, the set of variable nodes $V^1 \bigcup V^2 \bigcup \{v\}$ has $14$ check nodes, violating the $7 \rightarrow 16$ condition.
\end{proof}
Hence, if a decision is made after the second iteration, the decoder is successful.

\textbf{Subgraph 3}: The variables in $V^1$ induce the subgraph shown in Fig. \ref{Figure4c}. At the end of the first iteration, \VC v11, \VC v12 and \VC v13 receive correct messages from all their neighboring check nodes. Moreover, there exists no variable which receives four incorrect messages from checks in the set $C^1$. Hence, if a decision is made after the first iteration, the decoder is successful.

\textbf{Subgraph 4}: The variables in $V^1$ induce the subgraph shown in Fig. \ref{Figure4d}. At the end of the first iteration:
\begin{eqnarray}
\mb cv1 & = & \left \{
\begin{array}{rl}
1 & \text{if }c \in C^1 \backslash \{ \VC c14, \VC c17 \},~v \notin V^1 \\
1 & \text{if }c \in \{ \VC c14, \VC c17 \},~v \in V^1\\
0 & \text{otherwise}
\end{array} \right.
\end{eqnarray}
For no $v \in V \backslash V^1$, $\mb :v1 = \{1\}$ as it this would introduce a four-cycle. For any $v \in V \backslash V^1$, $\mf vc2 = 1$ only if $\mb {:\backslash c}v1 = \{1\}$. This implies that $v$ has three checks in the set $C^1 \backslash \{\VC c14, \VC c17 \}$. Let $V^2$ be the set of such variables. We now have the following lemma:
\begin{lema}
There can be at most two variables in $V^2$. Moreover, there exists no check $c \in C \backslash C^1$ which is shared by two variables in the set $V^2$.
\end{lema}
\begin{proof}
Let $\VC v21, \VC v22, \VC v23 \in V^2$. Then, the set $V^1 \bigcup \{ \VC v21,\VC v22, \VC v23 \}$ has at most $13$ checks which violates the $ 6 \rightarrow 14$ condition. Hence, $V^2$ has at most two variables.

Next, let two variables $\VC v21, \VC v22 \in V^2$ share a check $c \in C \backslash C^1$. Then, the set $V^1 \bigcup \{\VC v21, \VC v22\}$ has at most 11 checks which violates the $5 \rightarrow 12$ condition.
\end{proof}
Let $C^2$ be the set of checks $C \backslash C^1$ which are connected to variables in $V^2$. In the second iteration we have:
\begin{eqnarray}
\mf vc2 & = & \left \{
\begin{array}{rl}
1 & \text{if }v \in \{ \VC v11, \VC v13 \},~c \notin \{ \VC c14, \VC c17 \} \\
1 & \text{if }v = \VC v12 \\
1 & \text{if }v \in V^2,~c \in C^2 \\
0 & \text{otherwise}
\end{array}
\right. \\
\mb cv2 & = & \left \{
\begin{array}{rl}
1 & \text{if }c \in \{\VC c14, \VC c17\},~v \neq \VC v12 \\
1 & \text{if }c \in C^1 \backslash \{\VC c14, \VC c17\},~v \notin V^1 \\
1 & \text{if }c \in C^2,~v \notin V^2 \\
0 & \text{otherwise}
\end{array}
\right.
\end{eqnarray}
For no $v \in V \backslash V^1$, $\mb:v2 = \{1\}$, for such a structure cannot exist without creating a four-cycle or violating one of $5 \rightarrow 12$ and $6 \rightarrow 14$ conditions. For any $v \in V \backslash \left ( V^1 \bigcup V^2 \right )$, $\mb vc3 = 1$ only if $\mb {: \backslash c}v2 = \{1\}$. This implies that $v$ has three neighbors in the set $C^1 \bigcup C^2$. Let $V^3$ be the set of such variables. We have the following lemma:
\begin{lema} For the sets $V^2$ and $V^3$, the following are true:\\
(i) If $\left |V^2 \right | = 2$, then $V^3$ is empty. \\
(ii) If $\left | V^3 \right | > 0$, then $\left | V^2 \right | = 1$. \\
(iii) $\left | V^3 \right | \leq 1$.
\end{lema}
\begin{proof}
We prove the lemma part by part.\\
(i) Suppose $\left |V^2 \right | = 2$ and that $V^3$ is not empty. Let $\VC v31 \in V^3$. Then, the set $V^1 \bigcup V^2 \bigcup \{\VC v31\}$ is of size 6 and has at most $13$ checks which violates the $6 \rightarrow 14$ condition. \\
(ii) Suppose $\left | V^3 \right | > 0$. Let $\VC v31 \in V^3$. If $V^2$ is empty, then \VC v31 is connected to three checks in $C^1$. This is not possible as $\VC v31 \in V \backslash \left ( V^1 \bigcup V^2 \right )$. Next, suppose that $\left | V^2 \right | = 2$. Let $\VC v21, \VC v22 \in V^2$. Then $V^1 \bigcup \{\VC v21, \VC v22, \VC v31 \}$ has at most $13$ checks which violates the $6 \rightarrow 14$ condition.\\
(iii) Suppose $\VC v31, \VC v32 \in V^3$. By (ii), there exists a variable $\VC v21 \in V^2$. Then, $V^1 \bigcup \{\VC v21, \VC v31, \VC v32\}$ has at most $13$ checks which violates the $6 \rightarrow 14$ condition.
\end{proof}
Suppose $V^3 = \{\VC v31 \}$. Denote the fourth check of $\VC v31$ by $\VC c31$. Then, we have at the beginning of the third iteration:
\begin{equation}
\mf vc3 = \left \{
\begin{array}{rl}
1 & \text{if }v \in \{ \VC v11, \VC v13\},~c \notin \{ \VC c14, \VC c17 \} \\
1 & \text{if }v \in V^2,~c \in C^2 \\
1 & \text{if }v = \VC v31,~c = \VC c31
\end{array}
\right.
\end{equation}
At the end of the fourth iteration, $\mb :v3 = \{0\}$ for $v \in V^1$. Thus, if a decision is made at the end of this iteration, all $v \in V^1$ are decoded correctly. Now we prove the following lemma:
\begin{lema}
There exists no $v \in V \backslash  V^1$ such that $\mb :v3 = \{1\}$.
\end{lema}
\begin{proof}
Suppose that $V^3$ is empty and that there exists a variable $v$ such that $\mb :v3 = \{1\}$. If $V^2$ is empty, then, $v$ is connected to four checks in $C^1 \backslash \{\VC c14, \VC c15, \VC c16, \VC c17\} $. This is not possible as it would cause a four-cycle. If $V^2$ is not empty, then $v$ is connected to four checks in $\left (C^1 \backslash \{\VC c14, \VC c15, \VC c16, \VC c17\} \right ) \bigcup C^2$. Then, we would have $\mb :v2 = \{1\}$. However, from above, no such variable exists. 

Next, suppose that $V^3 = \{\VC v31\}$ and that $C^3 = \{ \VC c31 \}$. Then, \VC c31 is the only check such that $\mb c:2 = \{0\}$ and $\mb c:3 = \{1\}$. It follows then that for any $v \in V \backslash V^1$, if $\mb :v3 = \{1\}$, then $v$ is connected to \VC c31. Also, it is connected to three checks in the set $\{\VC c11, \VC c12, \VC c13, \VC c18, \VC c19, \VC c1{10}, \VC c21 \}$. Then the set of variables $V^1 \bigcup \{ \VC v21, \VC v31, v\}$ has at most $12$ checks. This violates the $6 \rightarrow 14$ condition.
\end{proof}
Hence, if a decision is made after the third iteration, the decoder is successful.

\textbf{Subgraph 5}: The variables in $V^1$ induce the subgraph shown in Fig. \ref{Figure4e}. For all $v \in V^1$, $~\mf v:i = \{1\},~1 \leq i \leq 3$. There exist no $v \in V \backslash V^1$ that receive three incorrect messages, for the existence of such a variable would violate the $4 \rightarrow 11$ condition. Hence, for all $v \in V \backslash V^1$, $\mf v:i = \{0\},~1 \leq i \leq 3$. 

Let $V^2$ be the set of variables that have two checks in the set $C^1 \backslash \{ \VC c11, \VC c14, \VC c17 \}$. Let $C^2$ be the remaining two checks of these variables. At the beginning of the fourth iteration, the decoder switches to the Gallager-B mode. Then:
\begin{equation}
\mf vc4 = \left \{
\begin{array}{rl}
1 & \text{if }v \in V^1,~c \in C^2 \backslash \{\VC c11, \VC c14, \VC c17\} \\
1 & \text{if }v \in V^2,~c \in C^2 \\
0 & \text{otherwise}
\end{array}
\right.
\end{equation}
At the end of the fourth iteration, $\mb :v4 = \{0\}$ for $v \in V^1$. Moreover, for no $v \in V \backslash V^1$, $\mb :v4 = \{1\}$, as the existence of such a variable would either induce a four-cycle or violate one of the $5 \rightarrow 12$, $6 \rightarrow 14$, $7 \rightarrow 16$ or $8 \rightarrow 18$ conditions (the arguments used are similar to the ones used for Subgraph 2 and Subgraph 4). Hence, if a decision is made at the end of the fourth iteration, the decoder is successful.

Now we prove the necessity of the conditions of the theorem. We prove this by giving subgraphs which violate \textit{one} condition and are not successfully decoded in four iterations. Since the validity of these claims can be checked easily, a detailed proof is omitted.

\textbf{Necessity of the $\mathbf{4 \rightarrow 11}$ condition}

Consider the subgraph shown in Fig. \ref{Figure4_10_Subgraph}. In this case, the $4 \rightarrow 11$ condition is not satisfied and the errors are not corrected at the end of the fourth iteration. Hence, in order to guarantee the correction of three errors in four iterations, the $4 \rightarrow 11$ condition must be satisfied.
\begin{figure}[!tbh]
\centering
\includegraphics[width=0.5\textwidth]{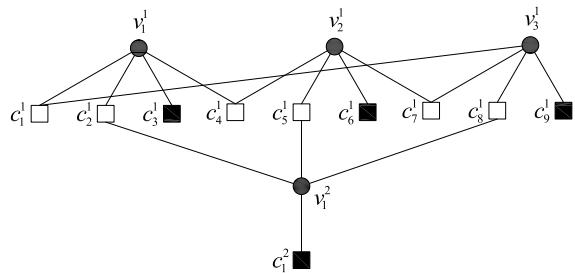}
\caption{A $4 \rightarrow 10$ subgraph\label{Figure4_10_Subgraph}}
\end{figure}

\textbf{Necessity of the $\mathbf{5 \rightarrow 12}$ condition}

There exists no graph of girth six which satisfies the $4 \rightarrow 11$ condition but does not satisfy the $5 \rightarrow 12$ condition. 

\textbf{Necessity of the $\mathbf{ 6 \rightarrow 14 }$ condition}

Consider the graph shown in Fig. \ref{Figure6_13_Subgraph}. The graph shown satisfies the $4 \rightarrow 11$ and the $5 \rightarrow 12$ conditions but not the $6 \rightarrow 14$ condition. The errors are not corrected in four iterations. Hence, in order to guarantee the correction of three errors in four iterations, the $6 \rightarrow 14$ condition must be satisfied.
\begin{figure}[!tbh]
\centering
\includegraphics[width=0.5\textwidth]{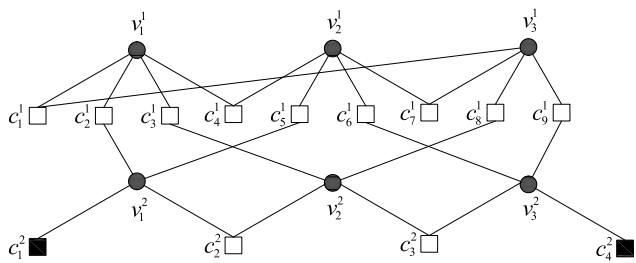}
\caption{A $6 \rightarrow 13$ subgraph\label{Figure6_13_Subgraph}}
\end{figure}

\textbf{Necessity of the $\mathbf{ 7 \rightarrow 16 }$ condition}

Consider the graph shown in Fig. \ref{Figure7_15_Subgraph}. The graph shown satisfies the probability $4 \rightarrow 11$, $5 \rightarrow 12$ and the $6 \rightarrow 14$ conditions but not the $7 \rightarrow 16$ condition. The errors are not corrected at the end of the fourth iteration. Hence, in order to guarantee the correction three errors in four iterations, the $7 \rightarrow 16$ condition must be satisfied.
\begin{figure}[!tbh]
\centering
\includegraphics[width=0.5\textwidth]{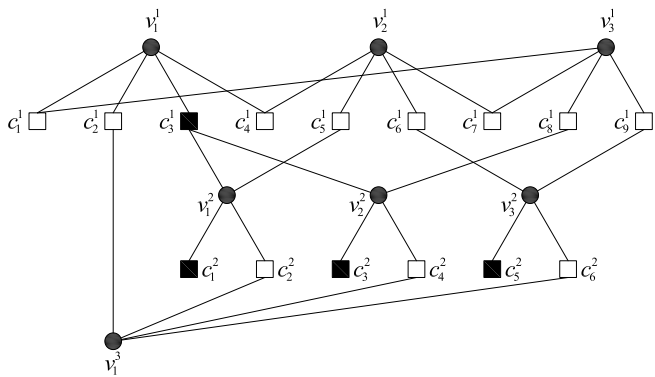}
\caption{A $7 \rightarrow 15$ subgraph\label{Figure7_15_Subgraph}}
\end{figure}

\textbf{Necessity of the $\mathbf{ 8 \rightarrow 18 }$ condition}

Consider the graph shown in Fig. \ref{Figure8_17_Subgraph}. The graph shown satisfies the $4 \rightarrow 11$, $5 \rightarrow 12$, $6 \rightarrow 14$ and the $7 \rightarrow 16$ condition but not the $8 \rightarrow 18$ condition. The errors are not corrected at the end of the fourth iteration. Hence, in order to guarantee the correction of three errors in four iterations, the $8 \rightarrow 18$ condition must be satisfied.
\begin{figure}[!tbh]
\centering
\includegraphics[width=0.5\textwidth]{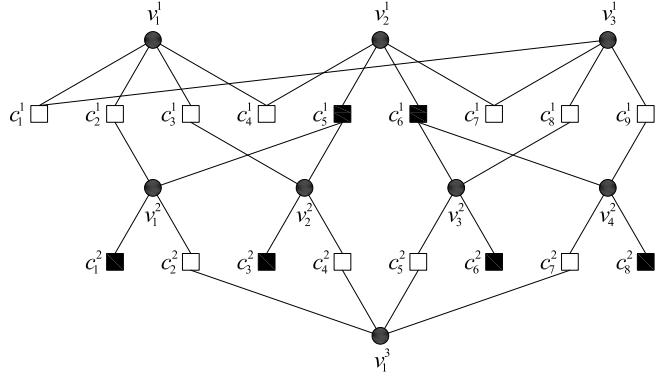}
\caption{A $8 \rightarrow 17$ subgraph\label{Figure8_17_Subgraph}}
\end{figure}
\end{proof}

In this section, we proved necessary and sufficient conditions to guarantee the correction of three errors in column-weight-four codes using an iterative decoding algorithm. By analyzing the messages being passed in subsequent iterations, it may be possible to get smaller bounds on the number of check nodes required in the ``small'' subgraphs. However, we hypothesize that the size of subgraphs to be avoided would be larger.

\section{Numerical Results}\label{section5}
In this section, we describe a technique to construct codes with column-weight three and four which can correct three errors. Codes capable of correcting a fixed number of errors show superior performance on the BSC at low values of transition probability $\alpha$. This is because the slope of the FER curve is related to the minimum critical number \cite{isitpaper}. A code which can correct $i$ errors has minimum critical number at least $i+1$ and the slope of the FER curve is $i+1$. We restate the arguments from \cite{isitpaper} to make this connection clear. 

Let $\alpha$ be the transition probability of a BSC and $c_k$ be the number of configurations of received bits for which $k$ channel  errors lead to codeword (frame) error. The frame error rate (FER) is given by:

$$FER(\alpha)=\sum_{k=i}^n c_k\alpha^k(1-\alpha)^{(n-k)}$$  where $i$ is the minimal number of channel errors that can lead to a decoding error and $n$ is length of the code.

On a semi-log scale the FER is given by
\begin{eqnarray}
\log \left ( FER(\alpha) \right )& = & \log \left ( \sum_{k=i}^n c_k\alpha^k(1-\alpha)^{n-k} \right )  \nonumber \\
& = & \log(c_i)+i\log(\alpha)+\log \left ( (1-\alpha)^{n-i} \right ) \nonumber \\
& & \quad +\log\left(1+\frac{c_{i+1}}{c_i}\alpha(1-\alpha)^{-1}+\ldots+\frac{c_{n}}{c_i}\alpha^{n-i}(1-\alpha)^{-i}\right) \nonumber
\label{other}
\end{eqnarray}

For small $\alpha$, the expression above is dominated by the first two terms. That is,
$$\log \left ( FER(\alpha) \right ) \approx \log(c_i)+i\log(\alpha)$$

The $\log(FER)$ vs. $\log(\alpha)$ graph is close to a straight line with slope equal to $i,$ the minimal critical number. If two codes $C_1$ and $C_2$ have minimum critical numbers $i_1$ and $i_2,$ such that $i_1 > i_2,$ then the code $C_1$ will perform better than $C_2$ for small enough $\alpha,$ independent of the number of trapping sets.

\subsection{Column-Weight-Three Codes}
From the discussion in above and in Section \ref{section3}, it is clear that for a code to have an FER curve with slope at least $4$, the corresponding Tanner graph should not contain the trapping sets shown in  Fig. \ref{trappingsets} as subgraphs. We now describe a method to construct such codes. The method can be seen as a modification of the PEG construction technique used by Hu \textit{et al.} \cite{peg}. The algorithm is detailed below as Algorithm \ref{Algorithm:1}.

\begin{algorithm}\label{Algorithm:1}
 \KwData{The set of $n$ variable nodes ($V$) and $m$ check nodes ($C$). The column weight of the code ($\gamma$)}
 \KwResult{ Code with column weight $\gamma$}
 \For{$j = 1$ to $n$}{
 		\For{$k = 1$ to $\gamma$}{
 				\eIf{$k = 1$}{
 						Connect the $k^{th}$ edge of variable node $j$ to the check node with the smallest positive degree.
 				}
 				{
 						Expand the tree rooted at node $j$ to a depth of $6$.\\
 						Assimilate all check nodes which do not appear in the tree into $C_{j,\overline{T}}$, the set of candidates for connecting variable node 								$j$ to.\\
 						\While{$k^{th}$ edge is not found}{
 								Find the check node $c_i$ in $C_{j,\overline{T}}$ with the lowest degree. If connecting $c_i$ to variable node $j$ does not create a 										$(5, 3)$ trapping set, set this as the $k^{th}$ edge. If it does, remove $c_i$ from $C_{j,\overline{T}}$.
 						}
 				}
 		}
 }
 \caption{ConstructCode}
\end{algorithm}
Note that checking for a graph isomorphic to $(8,0)$ trapping set is computationally complex. Since, the PEG construction empirically gives good codes, it is unlikely that it introduces a weight-eight codeword. However, once the graph is grown fully, it can be checked for the presence of weight-eight codewords and these can be removed by swapping few edges. 

\begin{figure}
\centering
\includegraphics[width=0.6\textwidth]{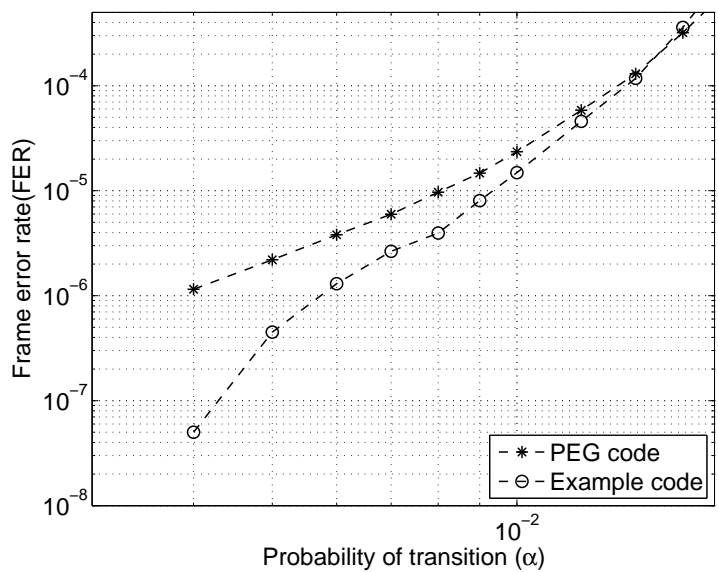}
\caption{Performance comparison of the PEG code and the example code for column-weight three\label{pegnewvsold}
}
\end{figure}

Using the above algorithm, a column-weight-three code with $504$ variable nodes and $252$ check nodes was constructed. The code has slight irregularity in check degree. There is one check node degree five and one check node with degree seven, but the remaining have degree six. The code has rate 0.5. In the algorithm, we restrict maximum check degree to seven. The performance of the code on BSC is compared with the PEG code of same length. The PEG code is empirically the best known code at that length on AWGN channel \cite{mackayswebsite}. However, it has fourteen $(5,3)$ trapping sets. Fig. \ref{pegnewvsold} shows the performance comparison of the two codes. As can be seen, the new code performs better than the original PEG code at small values of $\alpha$.

\subsection{Column-Weight-Four Codes}
\begin{figure}[htb]
\centering
\subfigure[] % caption for subfigure a
{\label{Figure:4_11_subgraph_1}\includegraphics[width=0.4\textwidth]{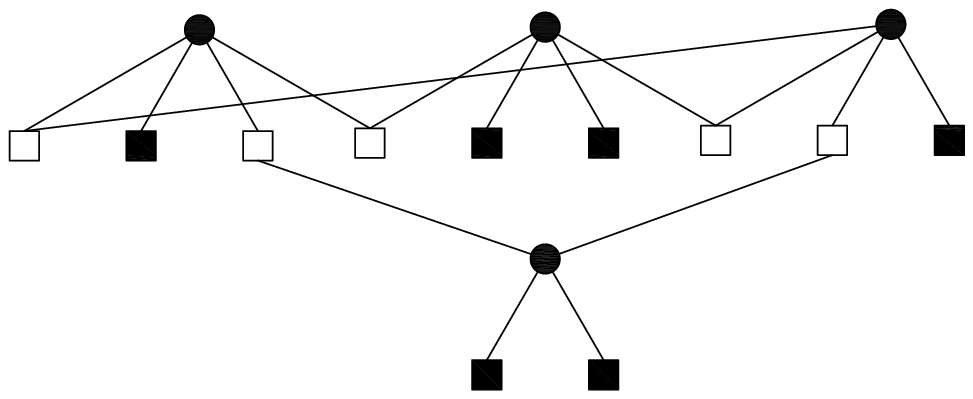} }
\hspace{0.05\textwidth}
\subfigure[] % caption for subfigure a
{\label{Figure:4_11_subgraph_2}\includegraphics[width=0.4\textwidth]{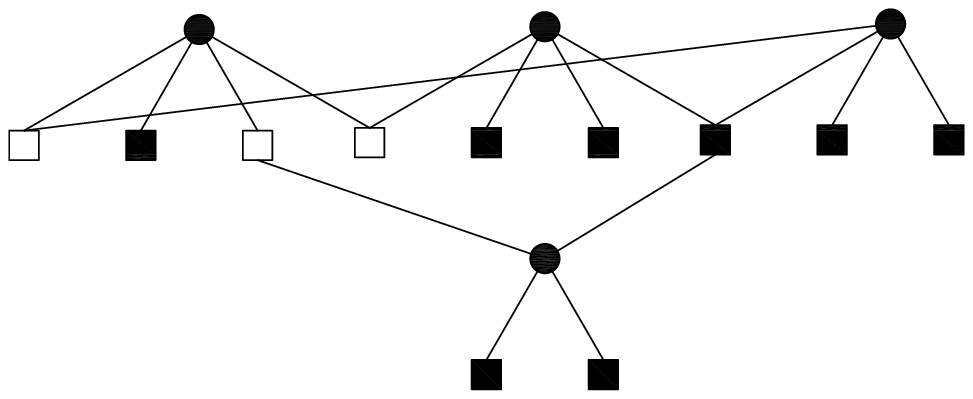} }
\caption{Graphs with girth 6 which have $4$ variable nodes and $11$ check nodes. Subgraphs with $4$ variable nodes and fewer than $11$ check nodes do not exist.\label{Figure:4_11} }
\end{figure}

Unlike column-weight-three codes, the construction of column-weight-four codes involves ensuring certain expansion on subsets of variable nodes. This can be done only in time which grows exponentially with the length of the code. Hence,we consider the $4 \rightarrow 12$ condition rather than the necessary and sufficient conditions discussed in Section \ref{section4}. It can be shown that the $4 \rightarrow 12$ condition is sufficient for the $4 \rightarrow 11$, $5 \rightarrow 12$, $6 \rightarrow 14$, $7 \rightarrow 16$ and the $8 \rightarrow 18$ conditions. There are only two graphs of girth $6$ with $4$ variable nodes and $11$ check nodes. Fig. \ref{Figure:4_11} shows these two graphs. Avoiding these two subgraphs will ensure a code which can correct three errors. An algorithm for the construction of such codes is similar to the modified PEG algorithm given in Algorithm \ref{Algorithm:1}. This algorithm was used to generate a code of length $816$, girth $6$ and rate $0.5$. The code constructed has a slight irregularity in that three check nodes have degree nine and three have degree seven.

\textit{Remark:} For the code parameters given above, it was possible to generate a code which satisfied the $4 \rightarrow 12$ condition. However, it might not be possible to satisfy this condition for codes with higher rate and/or shorter lengths. Should such a scenario arise, the set of subgraphs to be avoided should be changed (\textit{e.g.}, to those specified in the necessary and sufficient conditions). However, the code construction time will be larger. Hence, at the cost of code-construction time and complexity, it is possible to achieve shorter lengths and/or higher rates.

\begin{figure}[htb]
\centering
\includegraphics[width=0.6\textwidth]{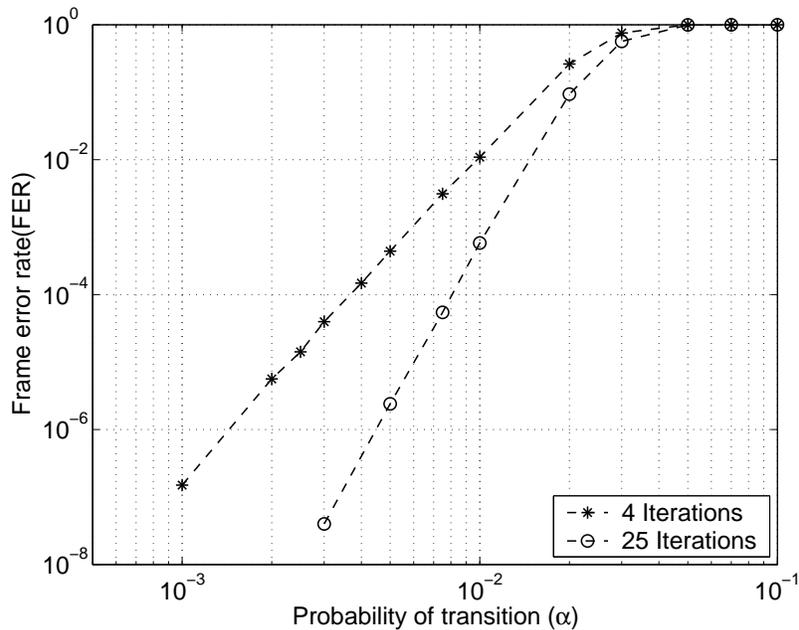}
\caption{Performance of the example column-weight-four code for different numbers of iterations of message-passing\label{Figure:ColWtFourPlot} }
\end{figure}

Fig. \ref{Figure:ColWtFourPlot} shows the performance of the code under message-passing decoding. The curve on the left corresponds to four iterations of message-passing. The curve in the right corresponds to $25$ iterations of message-passing.

After only four iterations, errors of weight four and above were encountered which were not corrected by the message-passing decoder. However, after $25$ iterations, the smallest weight error pattern still remaining had a weight of $7$. We note that the average slope of the FER curve is $8$ which is the weight of the dominant error event at these probabilities of error. This suggests that analysis over a higher number of iterations and on ``larger'' subgraph search will yield a stronger result. However, this is beyond the scope of this paper. Also, it is worth noting that the conditions of Theorem \ref{Theorem:4:1} avoid codewords of length $4$ through $8$ which improves the minimum distance of the code.

\section{Conclusion}\label{section6}
In this paper, we provided a method to derive conditions that guarantee the correction of a finite number of errors by hard-decision decoding. Although more involved than the expander arguments used in previous works, it results in better bounds. Moreover, in contrast to previous expansion arguments, our results give rise to code-construction techniques that yield codes with guaranteed error-correction ability under message-massing decoding at \textit{practically feasible lengths}. This method can be applied to (a) provide conditions for guaranteed correction of a larger number of errors, (b) yield similar results for higher column-weights and/or higher girths. However, such applications would be more involved than the analysis done in this work.
\bibliographystyle{IEEEtran}
%\bibliography{ref2}
% Generated by IEEEtran.bst, version: 1.12 (2007/01/11)

\end{document}